\documentclass[preprint,12pt]{emulateapj}

\usepackage{natbib}
\usepackage[T1]{fontenc}
\usepackage{graphicx}
\usepackage[space]{grffile}
\usepackage{latexsym}
\usepackage{amsfonts,amsmath,amssymb}

\newcommand{\kepmag}{\textit{Kp}\xspace}

\usepackage{xspace}
\usepackage{amsmath}

\newcommand{\Mstar}{\ensuremath{M_{\star}}\xspace}
\newcommand{\Rstar}{\ensuremath{R_{\star}}\xspace} 
\newcommand{\Lstar}{\ensuremath{L_{\star}}\xspace} 
\newcommand{\Rsun}{\ensuremath{R_{\odot}}\xspace }
\newcommand{\Msun}{\ensuremath{M_{\odot}}\xspace}
\newcommand{\Lsun}{\ensuremath{L_{\odot}}\xspace} 
\newcommand{\teff}{T$_{\mathrm{eff}}$\xspace}
\newcommand{\teq}{T$_{\mathrm{eq}}$\xspace}
\newcommand{\gcc}{g~cm$^{-3}$\xspace}

\def\deg{\ensuremath{^{\circ}}}

\newcommand{\RH}{\ensuremath{R_H}}
\newcommand{\Min}{\ensuremath{M_\mathrm{in}}}
\newcommand{\Mout}{\ensuremath{M_\mathrm{out}}}
\newcommand{\ain}{\ensuremath{a_\mathrm{in}}}
\newcommand{\aout}{\ensuremath{a_\mathrm{out}}}

\newcommand{\Per}{\ensuremath{P}\xspace}
\renewcommand{\Re}{\ensuremath{R_{\oplus}}\xspace} 
\newcommand{\Me}{\ensuremath{M_{\oplus}}\xspace} 
\newcommand{\Mp}{\ensuremath{M_{P}}\xspace} 
\newcommand{\Rp}{\ensuremath{R_P}\xspace}
\newcommand{\Sinc}{\ensuremath{S_\mathrm{inc}}\xspace}
\newcommand{\Se}{\ensuremath{S_{\oplus}}\xspace}
\newcommand{\um}{\ensuremath{\mu \mathrm{m}}\xspace}

\newcommand{\Kepler}{\textit{Kepler}\xspace} 
\newcommand{\ktwo}{\textit{K2}\xspace}

\newcommand{\ms}{m s$^{-1}$\xspace}

\newcommand{\Rpb}{1.59 $\pm$ 0.43\xspace}
\newcommand{\Rpc}{1.92 $\pm$ 0.53\xspace}
\newcommand{\Pb}{9.32414\xspace}
\newcommand{\Pc}{15.50120\xspace}
\newcommand{\Prat}{1.6624\xspace}
\newcommand{\epic}{EPIC-206011691\xspace}
\newcommand{\ktwoid}{K2-21\xspace}
\newcommand{\ktwoidb}{K2-21b\xspace}
\newcommand{\Mb}{4\xspace}
\newcommand{\kb}{1.7\xspace}
\newcommand{\Mc}{5\xspace}
\newcommand{\kc}{2.0\xspace}
\newcommand{\fehepic}{\ensuremath{-0.11 \pm 0.13}\xspace}
\newcommand{\Rstarepic}{\ensuremath{0.60  \pm 0.10}\xspace}
\newcommand{\Lstarepic}{\ensuremath{0.086 \pm 0.064}\xspace}
\newcommand{\Mstarepic}{\ensuremath{0.64  \pm 0.11}\xspace}
\newcommand{\teffepic}{\ensuremath{4043 \pm 375}\xspace}
\newcommand{\RRatb}{\ensuremath{2.60 \pm 0.14}}
\newcommand{\RRatc}{\ensuremath{3.15 \pm 0.20}}
\newcommand{\Ksmagepic}{\ensuremath{9.42 \pm 0.02}\xspace}
\newcommand{\distepic}{\ensuremath{65 \pm 6}\xspace}
\newcommand{\teqepicb}{\ensuremath{510_{-70}^{+120}}\xspace}
\newcommand{\teqepicc}{\ensuremath{430_{-60}^{+100}}\xspace}
\newcommand{\rhostarcircb}{$2.7^{+3.7}_{-1.6}$} 
\newcommand{\rhostarcircc}{$5.4^{+5.4}_{-3.5}$} 
\newcommand{\rhostarspex}{$4.15 \pm 0.61$} 
\newcommand{\rhostar}{\ensuremath{\rho_\star}\xspace}
\newcommand{\rhostarcirc}{\ensuremath{\rho_{\star,\mathrm{circ}}}\xspace}

\shorttitle{Two Transiting Earth-size Planets Orbiting a Nearby M star}
\shortauthors{Petigura et al.}

\begin{document}


\title{Two Transiting Earth-size Planets Near Resonance Orbiting a Nearby Cool Star}


\author{
Erik A. Petigura\altaffilmark{1,10},
Joshua E. Schlieder\altaffilmark{2,11},
Ian J. M. Crossfield\altaffilmark{3,12},
Andrew W. Howard\altaffilmark{4},
Katherine M. Deck\altaffilmark{1},
David R. Ciardi\altaffilmark{5},
Evan Sinukoff\altaffilmark{4},
Katelyn N. Allers\altaffilmark{6,13},
William M. J. Best\altaffilmark{4},
Michael C. Liu\altaffilmark{4},
Charles A. Beichman\altaffilmark{1},
Howard Isaacson\altaffilmark{7},
Brad M. S. Hansen\altaffilmark{8}, 
S\'ebastien L\'epine\altaffilmark{9}
}

\altaffiltext{1}{California Institute of Technology, Pasadena, California, U.S.A. \href{mailto:petigura@caltech.edu}{petigura@caltech.edu}}
\altaffiltext{2}{NASA Ames Research Center, Moffett Field, CA, USA}
\altaffiltext{3}{Lunar \& Planetary Laboratory, University of Arizona, 1629 E. University Blvd., Tucson, AZ, USA}
\altaffiltext{4}{Institute for Astronomy, University of Hawaii, 2680 Woodlawn Drive, Honolulu, HI, USA}
\altaffiltext{5}{NASA Exoplanet Science Institute, California Institute of Technology, 770 S. Wilson Ave., Pasadena, CA, USA}
\altaffiltext{6}{Department of Physics and Astronomy, Bucknell University, Lewisburg, PA 17837, USA}
\altaffiltext{7}{Astronomy Department, University of California, Berkeley, CA, USA}
\altaffiltext{8}{Department of Physics \& Astronomy, University of California Los Angeles, Los Angeles, CA, USA}
\altaffiltext{9}{Department of Physics \&\ Astronomy, Georgia State University, Atlanta, GA, USA}
\altaffiltext{10}{Hubble Fellow}
\altaffiltext{11}{NASA Postdoctoral Program Fellow}
\altaffiltext{12}{NASA Sagan Fellow}
\altaffiltext{13}{Visiting Astronomer at the Infrared Telescope Facility, which is operated by the University of Hawaii under contract NNH14CK55B with the National Aeronautics and Space Administration}





\begin{abstract}
Discoveries from the prime \Kepler mission demonstrated that small planets (< 3~\Re) are common outcomes of planet formation. While \Kepler detected many such planets, all but a handful orbit faint, distant stars and are not amenable to precise follow up measurements. 
Here, we report the discovery of two small planets transiting \ktwoid, a bright ($K = 9.4$) M0 dwarf located \distepic~pc from Earth. We detected the transiting planets in photometry collected during Campaign 3 of NASA's \ktwo mission. Analysis of transit light curves reveals that the planets have small radii compared to their host star, \Rp/\Rstar = \RRatb\% and \RRatc\%, respectively. We obtained follow up NIR spectroscopy of \ktwoid to constrain host star properties, which imply planet sizes of \Rpb~\Re and \Rpc~\Re, respectively, straddling the boundary between high-density, rocky planets and low-density planets with thick gaseous envelopes. The planets have orbital periods of \Pb~days and \Pc~days, respectively, and a period ratio $P_c/P_b$~=~\Prat, very near to the 5:3 mean motion resonance, which may be a record of the system's formation history. Transit timing variations (TTVs) due to gravitational interactions between the planets may be detectable using ground-based telescopes. Finally, this system offers a convenient laboratory for studying the bulk composition and atmospheric properties of small planets with low equilibrium temperatures.
\end{abstract}


\bibliographystyle{apj}

\section{Introduction}
Analysis of photometry collected during the prime \Kepler mission (2009--2013) demonstrated that small planets are common around G, K, \& M dwarfs \citep{howard:2012,fressin:2013,petigura:2013b,dressing:2013}. 
Planets that are nearly Earth-size (1--2~\Re) and orbit close to their host stars (\Per~<~100~days) are abundant around Sun-like stars \citep[26\% of G \& K dwarfs host such a planet;][]{petigura:2013b} and nearly ubiquitious around M dwarfs \citep[1.2 planets per star;][]{dressing:2015b}. The higher occurrence of such planets around M dwarf hosts cannot be explained by higher detection efficiency for small planets around M dwarfs. Both \cite{petigura:2013b} and \cite{dressing:2015b} empirically derived survey completeness by injecting mock transits into \Kepler photometry and measuring the recovery rate. Both groups corrected for modest pipeline incompleteness ($\lesssim 50\%$ for $P$~<~100~days, \Rp~=~1--2~\Re). An open question is whether small planets are intrinsically more numerous around M dwarfs, or if M dwarf systems are more compact.

M dwarfs offer a convenient laboratory to study the bulk physical properties and atmospheres of small planets. 
Planet transits are deeper and radial velocity signatures are larger, for a given planet size and orbital period. 
Planets around M dwarfs having low equilibrium temperatures (e.g. \teq~=~200--600~K) have tighter orbits with shorter orbital periods compared to similar planets around solar-type stars.
These planets are especially compelling targets for atmospheric transmission spectroscopy by the James Webb Space Telescope (JWST), 
provided a sample of nearby, bright M dwarfs with warm transiting planets in the $\sim$1--3~\Re size range can be identified \citep{batalha:2015}.

Finding planets that transit M dwarfs is difficult because the observable stars are generally faint, and the brightest (nearest) ones are sparsely distributed on the sky.  
\Kepler surveyed a few thousand M dwarfs and only discovered $\approx$160 planet candidates, despite the high planet occurrence rate.
To capitalize on the follow up opportunities, bright stars are required.  
Although the \Kepler sample of M dwarfs was faint (median \kepmag~=~15.5~mag), the magnitude-limited brightness distribution naturally included some brighter targets.    
Among the most favorable is Kepler-138 (\kepmag~=~12.9 mag, $K$~=~9.5~mag), a system of three transiting planets, including two nearly Earth-size planets (1.2 \Re) in warm orbits \citep{Hutter15,kipping:2014}. 

Now that \Kepler is operating in the \ktwo mode, we have the opportunity to greatly expand the number of known small planets transiting bright M dwarfs. During \ktwo observations, the {\em Kepler Space Telescope} observes different regions of the ecliptic every 90 days, casting a wider net for nearby transiting planets \citep{howell:2014}.  Essentially, each of the planned fourteen \ktwo pointings offers the possibility of discovering a sample of short-period planets orbiting bright stars that is similar to the ensemble from the prime \Kepler mission.  

Here we present the discovery of a new multi-planet system orbiting a bright M dwarf in the \ktwo Campaign 3 field, \ktwoid, also known as \epic.  This system is a cousin of Kepler-138, identified during the prime \Kepler mission, and K2-3, a system of three transiting super-Earths discovered by our team in an earlier \ktwo Campaign \citep{crossfield:2015}.  
It also offers a preview of expected results from the TESS mission, when sky coverage will increase by another order of magnitude \citep{ricker:2015}.

In this paper, we describe our detection of \ktwoidb\ and c from \ktwo photometry along with our spectroscopic and imaging follow up in Section~\ref{sec:obs}. In Section~\ref{sec:analysis}, we present our analysis of stellar properties, planet properties, and false positive probability. Finally, we place the \ktwoid system in the context of other transiting planets in Section~\ref{sec:conclusions} and offer some thoughts regarding the rich follow up potential for this system.

\section{Observations}
\label{sec:obs}

\subsection{\ktwo Photometry}
\ktwoid was observed during \ktwo Campaign 3 lasting from 2014~November 14 to 2015 February 3. The star is listed as \epic in the Mikulski Archive for Space Telescopes (MAST). Target properties, including optical and NIR photometry from APASS \citep{henden:2012} and 2MASS \citep{skrutskie:2006} are listed in Table~\ref{tab:obs}.

\begin{figure*}[ht!]
\begin{center}
\includegraphics[width=7in]{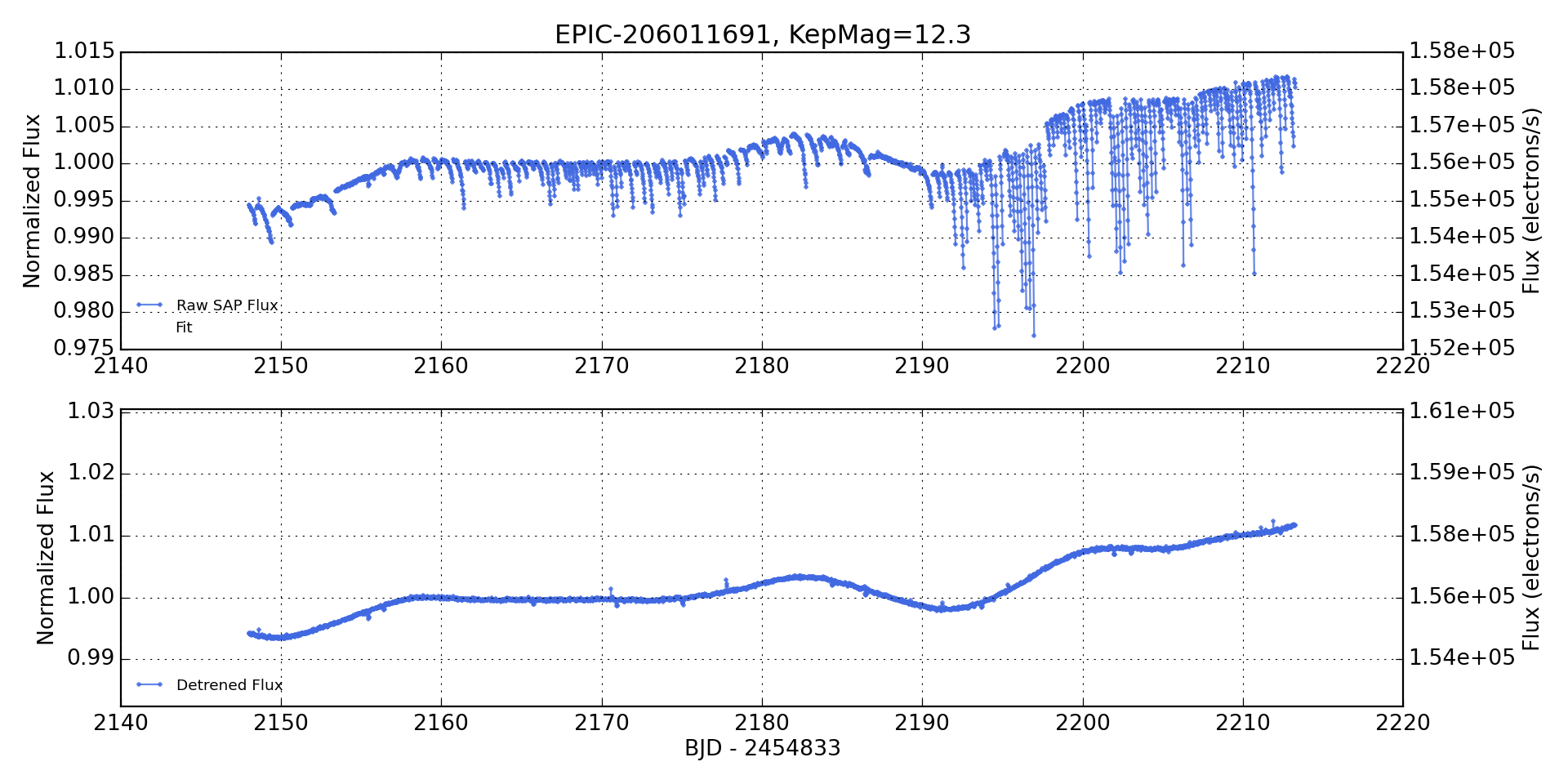}
\caption{{\em Top}: Raw photometry computed by summing the background-subtracted counts inside a circular aperture (3  pixel radius) centered on \ktwoid. {\em Bottom}: Photometry after correcting for variations due to telescope roll angle. Noise on three-hour timescales has been reduced by a factor of 30. The $\sim$1\% variability gives an upper limit to \ktwoid's intrinsic variability. However, since stars drift perpendicular to the roll direction over the course of a campaign, it is difficult to disentangle long-term astrophysical variability from position-dependent variability. The data used to produce the bottom panel is included as an electronic supplement.}
\label{fig:photometry}
\end{center}
\end{figure*}

We extracted \ktwo photometry for the star \ktwoid from the \Kepler pixel data, which we downloaded from the MAST. Our photometric extraction routine is outlined in \cite{crossfield:2015}. In brief, during \ktwo observations, stars drift across the CCD by $\sim$1 pixel every $\sim$6 hours. As the stars drift through pixel-phase, {\bf intra-pixel} sensitivity variations and errors in the flatfield cause the apparent brightness of the target star to change. For every observation we solve for the roll angle between the target frame and an arbitrary reference observation. We model the time- and roll-dependent brightness variations using a Gaussian process. Figure~\ref{fig:photometry} shows both the raw and corrected photometry. We do not see any evidence for stellar variability with timescales of <~5~days, suggesting that \ktwoid is a slowly rotating star. The corrected light curve 
exhibits $\sim$1\% variability on timescales longer than $\sim$10~days. This provides an upper limit on the intrinsic variability of \ktwoid over long timescales. However, since stars drift perpendicular to the roll direction over the course of a campaign, it is difficult to disentangle long-term astrophysical variability from position-dependent variability. Our calibrated \ktwo photometry is included as an electronic supplement.

{\renewcommand{\arraystretch}{1.2}
\hspace{-1in}
\begin{deluxetable*}{l l l l }[bt]
\tabletypesize{\scriptsize}
\tablecaption{Stellar Parameters of \ktwoid \label{tab:obs}}
\tablewidth{0pt}
\tablehead{
\colhead{Parameter} & Units & \colhead{Value} & \colhead{Source}
}
\startdata
\multicolumn{3}{l}{\hspace{1cm}Identifying information} \\
EPIC ID       & --      & 206011691                    & EPIC \\
2MASS ID      & --      & 22411288-1429202             & 2MASS  \\
$\alpha$ R.A. & h:m:s   & 22:41:12.89                  & EPIC \\
$\delta$ Dec. & d:m:s   & $-$14:29:20.35               & EPIC \\
$l$           & d:m:s   & 48:56:11.39                  & EPIC \\
$b$           & d:m:s   & $-$57:10:44.78               & EPIC \\
\multicolumn{3}{l}{\hspace{1cm}Photometric Properties} \\
\kepmag       & mag     & 12.31            & EPIC \\  
B             & mag     & 14.14 $\pm$ 0.06 & APASS  \\
V             & mag     & 12.85 $\pm$ 0.02 & APASS \\
g$^\prime$    & mag     & 13.53 $\pm$ 0.02 & APASS \\
r$^\prime$    & mag     & 12.32 $\pm$  0.06 & APASS \\
i$^\prime$    & mag     & 11.76 $\pm$ 0.06 & APASS \\
J             & mag     & 10.25 $\pm$ 0.02 & 2MASS\\
H             & mag     & 9.63  $\pm$  0.02 & 2MASS\\
Ks            & mag     & \Ksmagepic & 2MASS\\
\multicolumn{3}{l}{\hspace{1cm}Spectroscopic and Derived Properties} \\
$\mu_{\alpha}$& mas~yr$^{-1}$ & $17.3 \pm 1.4 $ & \cite{zacharias:2012} \\
$\mu_{\delta}$& mas~yr$^{-1}$ & $-78 \pm 2.8 $ & \cite{zacharias:2012} \\
Spectral Type & --      & M0.0$\pm$0.5  & SpeX, this paper\\
\teff         & K       & \teffepic     & SpeX, this paper\\
$[$Fe/H$]$    & dex     & \fehepic      & SpeX, this paper\\
\Mstar        & \Msun   & \Mstarepic    & SpeX, this paper\\
\Rstar        & \Rsun   & \Rstarepic    & SpeX, this paper\\
\rhostar      & \gcc    & \rhostarspex  & SpeX, this paper\\
\Lstar        & \Lsun   & \Lstarepic    & SpeX, this paper\\
Distance      & pc      & \distepic     & this paper\\
Age           & Gyr     & $\gtrsim$1    & this paper 
\enddata
\end{deluxetable*}
}

\subsection{Transit Detection}
\begin{figure*}[ht!]
\begin{center}
\includegraphics[width=7in]{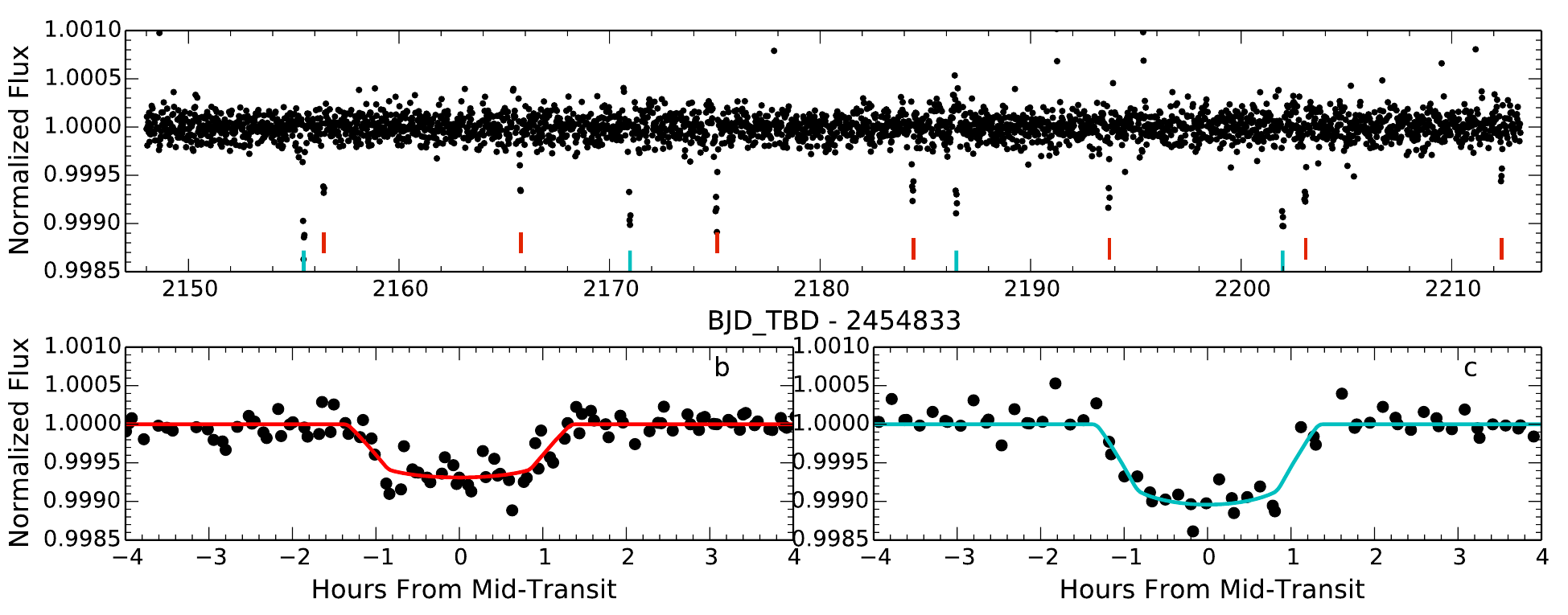}
\caption{Calibrated \ktwo photometry for \ktwoid.  Vertical ticks indicate the locations of each planet's transits. {\em Bottom}: Phase-folded photometry and best-fit light curves for each planet. Best fit parameters from light curve fitting are tabulated in Table~\ref{tab:planet}.}
\label{fig:photometry-fits}
\end{center}
\end{figure*}

We searched through the calibrated and detrended photometry 
(top panel of Figure~\ref{fig:photometry-fits}) using the TERRA algorithm described in \cite{petigura:2013a}. TERRA identified a transit candidate having
$P$~=~\Pc~days and signal-to-noise ratio (SNR)~=~25. We fit this candidate with a
\cite{mandel:2002} model and subtracted the best fit model from the
photometry. We reran TERRA on the photometry with the $P$~=~\Pc~day candidate removed. We found a second candidate having $P$~=~\Pb~days and SNR~=~22. Again we removed the best-fitting model. In subsequent searches, TERRA did not find any additional periodic box-shaped signals.

\subsection{Infrared and Optical Spectroscopy}
\label{sec:spectrsocopy}
We observed \ktwoid on 2015 July 23 UT using the recently refurbished SpeX spectrograph \citep{rayner:2003} on the 3.0~m NASA Infrared Telescope Facility (IRTF). The data were taken under clear skies with an average K-band seeing of 0.4--0.5 arcsec. We observed with the instrument in short cross dispersed mode (SXD) using the 0.3 $\times$ 15~arcsec slit. This setup provides simultaneous wavelength coverage from 0.7 to 2.5 $\mu$m at a resolution of R~$\approx$~2000. The extended blue wavelength coverage is a result of the recent chip upgrade SpeX received in  2014 July. The target was placed at two positions along the slit and observed in an ABBA pattern for subsequent sky subtraction. The observing sequence consisted of 8 $\times$ 60\,s exposures for a total integration time of 480\,s. Once the exposures were stacked, this integration time led to a signal-to-noise of $\sim200$ per resolution element. We obtained standard SpeX calibration frames consisting of flats and arclamp exposures immediately after observing \ktwoid.

\begin{figure*}
\begin{center}
\includegraphics[width=7in]{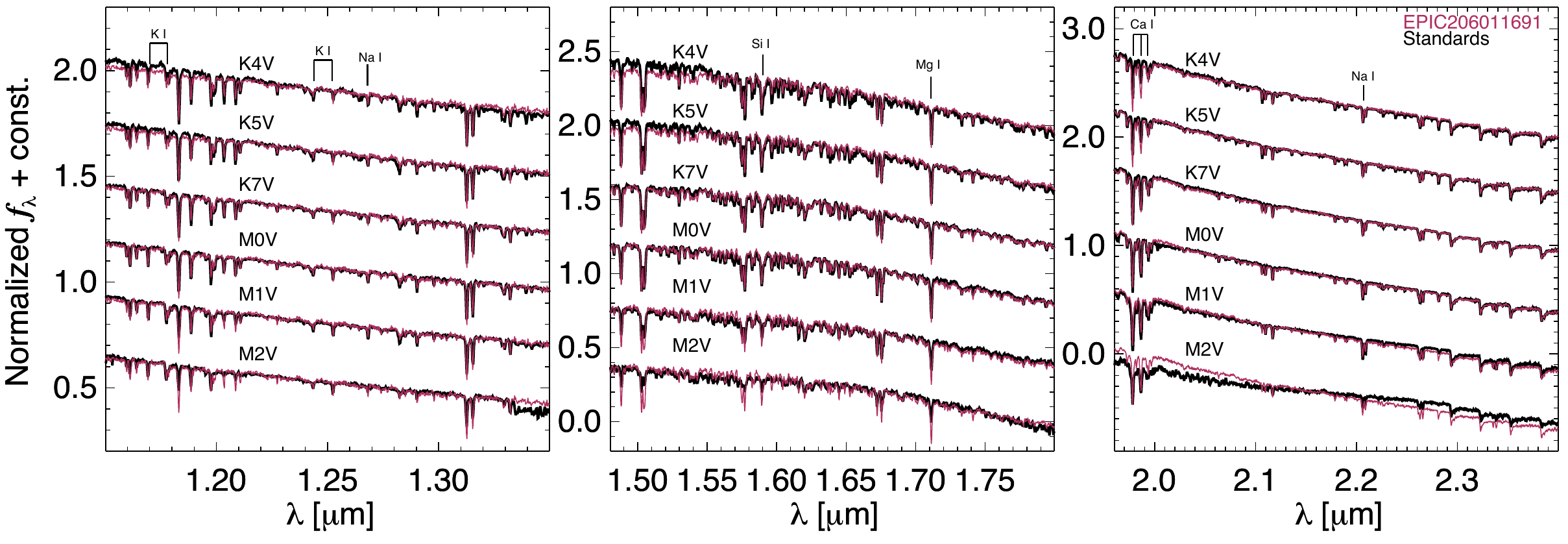}
\caption{Calibrated IRTF/SpeX spectra of \ktwoid compared to spectral standards. After fitting several spectral indices (described in Section~\ref{sec:stellar-parameters}), we derive a best-fit spectral type of M0.0$\pm$0.5. Spectroscopically derived stellar parameters are listed in Table~\ref{tab:obs}.}
\label{fig:spec} 
\end{center}
\end{figure*}

The SpeX spectrum was reduced using the SpeXTool package \citep{vacca:2003, cushing:2004}.
SpeXTool performs flat fielding, bad pixel removal, wavelength calibration, 
sky subtraction, spectral extraction and combination, telluric correction, 
flux calibration, and order merging. Flux
calibration and telluric corrections were perfomed using the spectrum of the
A0V-type star HD~218639 which was observed within 16 minutes and 0.04 airmass of
the target. The calibrated spectrum is compared to late-type standards from the 
IRTF Spectral Library \citep{cushing:2005, rayner:2009} in Figure~\ref{fig:spec}.
\ktwoid is a best visual match to the K7/M0 standards across the near-IR
bands. We perform a more detailed spectral typing in Section~\ref{sec:stellar-parameters}.

\ktwoid was also observed using Keck/HIRES \citep{vogt:1994} on 2015 July 24 UT with a total integration time of 216s. HIRES provides wavelength coverage from $\sim$3600-8000~\AA~at $R\approx60000$. The spectrum was reduced using the standard pipeline of the California Planet Search \citep{marcy:2008} and has SNR~$\approx$~45/pixel.

\subsection{Archival and Adaptive Optics Imaging}
\label{sec:ao}

\begin{figure}
\centering
\includegraphics[width=0.4\textwidth]{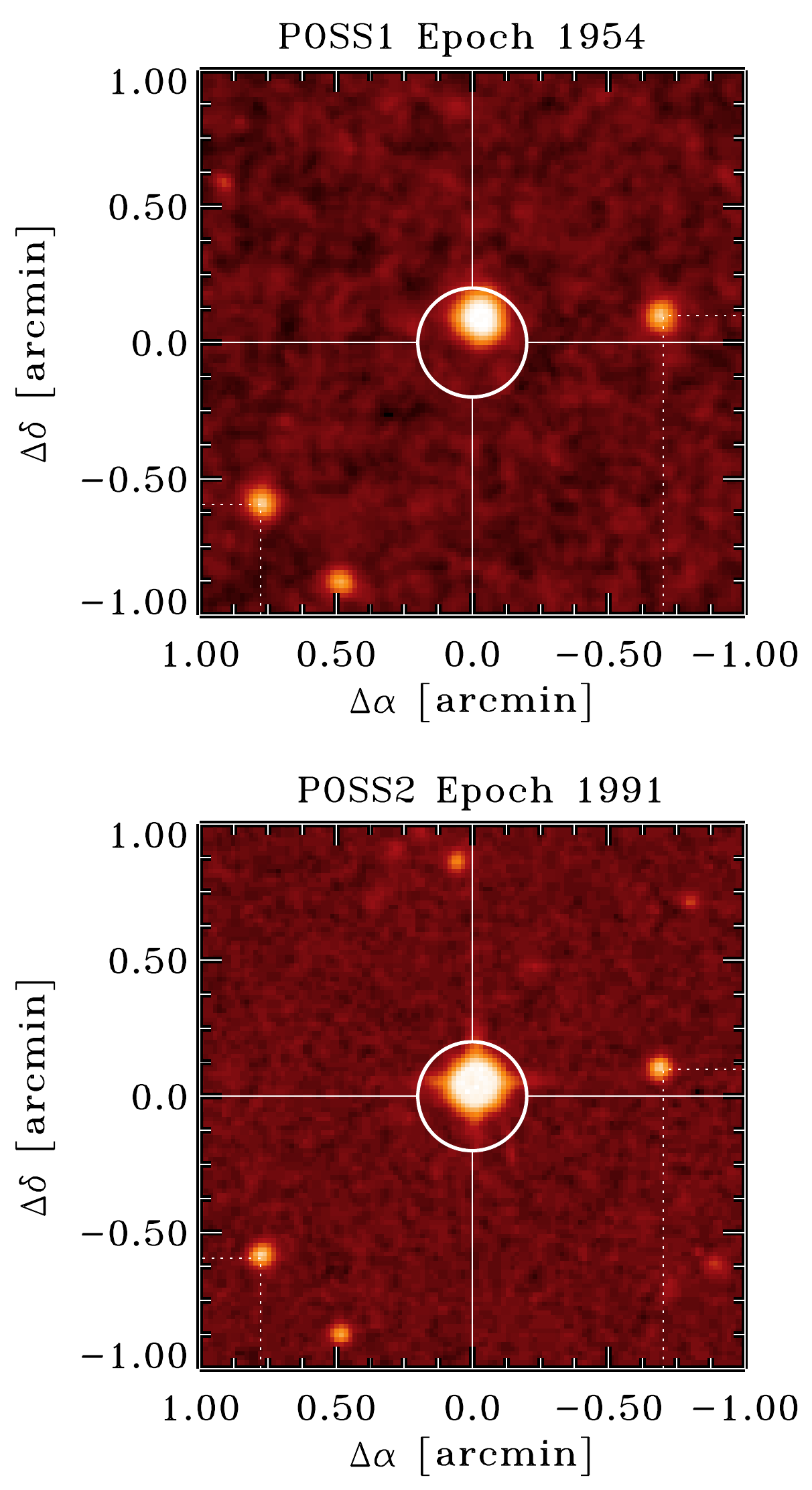}
\caption{POSS1 blue plates observed in 1954 (top panel) and POSS2 red plates observed in 1991 (bottom panel). The circle shows the location and extent of the circular aperture used to extract \ktwoid photometry at the 2015 position of the star. Between 1954 and 1991, \ktwoid moved by $\sim$3~arcsec, which can be clearly seen in the DSS images. In the POSS1 plate, the star is offset from the 2015 position by $\sim5$~arcsec. The DSS blue plates rule out a background star coincident with the current location of \ktwoid to $\Delta B =$ 2.5--3.0.}
\label{fig:dss} 
\end{figure}

To assess the possibility of background stars falling inside the software aperture of \Kepler pixels, we compare two epochs of imaging data from the Digitized Sky Survey (DSS). In Figure~\ref{fig:dss}, we show the DSS-Blue plates taken on 1954 July 25 (top panel) and the DSS-Red plates taken on 1991 July 17 (bottom panel). The images are 2$\times$2~arcmin and have a pixel scale of 1~arcsec/pixel. The images are centered on the epoch 2015 coordinates of the target as observed by \Kepler ($\alpha$: 22:41:12.9,\ $\delta$: -14:29:21.9 J2000.0) and the open circle represents the aperture size used when extracting the calibrated photometry.

The DSS images clearly show the proper motion of the primary target, while the nearby stars, located 40\arcsec\ W and 60\arcsec\ SE, show no significant astrometric motion. The primary target, in contrast, displays a clear proper motion of $\sim$3\arcsec\ over 37 years, in reasonable agreement with the measured proper motion \citep{lepine:2011,zacharias:2012}. In the DSS image there is no evidence of a background star at the 2015 position of \ktwoid. We estimate that if a star is located this position, that star must be at least 2.5--3 mag fainter than \ktwoid.  

We also obtained near-infrared adaptive optics images of EPIC 206011691 at Keck Observatory on the night of 2015 July 24 UT. Observations were obtained with the 1024 $\times$ 1024 NIRC2 array and the natural guide star system; the target star was bright enough to be used as the guide star. The data were acquired using the Kcont filter using the narrow camera field of view with a pixel scale of 9.942 mas/pixel. The Kcont filter has a narrower bandwidth (2.25--2.32~\um) compared the K filter (1.99--2.40~\um) and allows for longer integration times before saturation. A 3-point dither pattern was utilized to avoid the noisier lower left quadrant of the NIRC2 array. The 3-point dither pattern was observed thee times with 10 coadds and a 1.5 second integration time for a total on-source exposure time of $3 \times 3 \times 10 \times 1.5$s~=~135~s.

\begin{figure}
\centering
\includegraphics[width=0.45\textwidth]{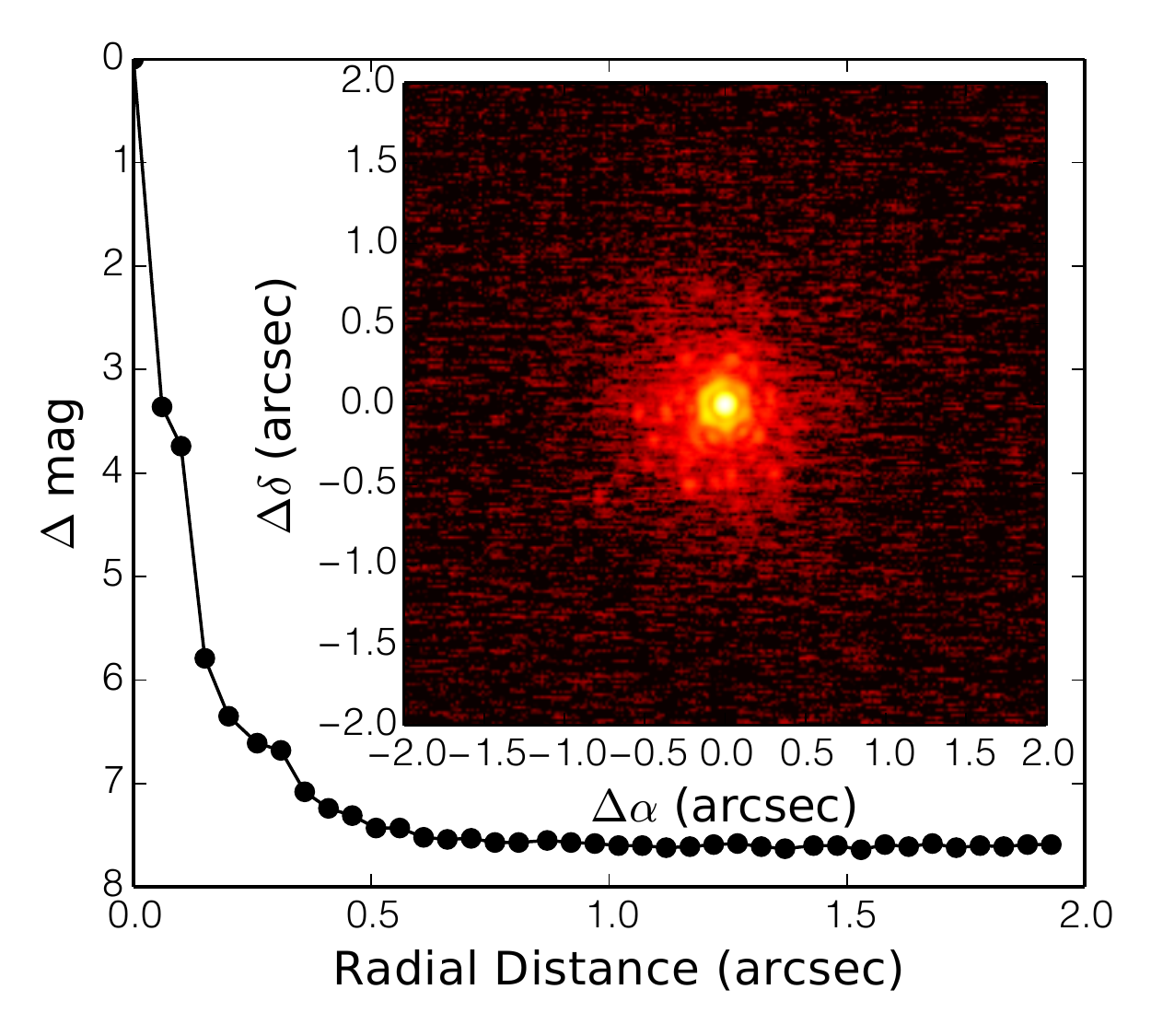}
\caption{NIRC2 K-band image and contrast curve. No stars with contrasts $\Delta K < 3.7$ are detected with separations > 0.1~arcsec and $\Delta K < 7.5$ with separations > 0.5~arcsec.}
\label{fig:keck} 
\end{figure}

The target star was measured with a resolution of 0.06~arcsec (FWHM). No other stars were detected within the 10~arcsec field of view of the camera. In the Kcont filter, the data are sensitive to stars that have K-band contrast of $\Delta K=3.7$ at a separation of 0.1~arcsec and $\Delta K=7.5$ at 0.5~arcsec from the central star. We estimate the sensitivities by injecting fake sources with a signal-to-noise ratio of 5 into the final combined images at distances of $N \times$ FWHM from the central source, where $N$ is an integer.  The 5$\sigma$ sensitivities, as a function of radius from the star, are shown in Figure~\ref{fig:keck}.

\section{Analysis}
\label{sec:analysis}

\subsection{Stellar Parameters}
\label{sec:stellar-parameters}
Here, we derive \ktwoid effective temperature, metallicity, mass, radius, and age from the IRTF/SpeX and Keck/HIRES spectra described in Section~\ref{sec:spectrsocopy}.
\cite{mann:2013a} provide spectral indices and empirical fits to estimate M dwarf
effective temperatures from optical and near-IR spectra. 
These indices and fits are calibrated using the M dwarf sample of
\cite{boyajian:2012b}. Following these methods, we estimate
the JHK-band temperatures of \ktwoid and calculate a mean \teff\ 
and rms scatter to find \teff = \teffepic~K ($\pm$148 K systematic error,
$\pm$375 K total error). This range of effective temperatures is consistent 
with main-sequence dwarfs spanning the transition between K- and M-types
\citep{pecaut:2013}. We use the metallicity calibration and
software\footnote{\url{https://github.com/awmann/metal}} of \cite{mann:2013b} 
to calculate \ktwoid's metallicity using the H- and K-bands. The final metallicity is
the mean of the H- and K-band estimates and the error is calculated from the quadrature
sum of the measurement error and systematic error in each band. We find [Fe/H] =
\fehepic~dex, approximately solar.
\cite{mann:2013a} also provide empirical calibrations and
software\footnote{\url{https://github.com/awmann/Teff\_rad\_mass\_lum}} 
to calculate the radii, masses, and luminosities of M dwarfs as a function of  \teff. Using the most conservative \teff\ errors, we calculate \Rstar~=~\Rstarepic~\Rsun, \Mstar~=~\Mstarepic~\Msun, and \Lstar~=~\Lstarepic~\Lsun.  These values, and the other fundamental parameters of the star, are tabulated in Table \ref{tab:obs} and are used for our calculations of the individual planet properties.


We also use our SpeX spectrum to estimate the spectral type of 
\ktwoid. We calculate the spectral type from several temperature
sensitive molecular indices across the spectrum. In the optical, 
the TiO5 and CaH3 indices
\citep{reid:1995,gizis:1997} are covered by the spectrum and are calibrated for the earliest M dwarfs
\citep{lepine:2003}. We follow the most recent spectral type
calibrations of these indices by \cite{lepine:2013} and estimate 
a spectral type of K7.5$\pm$0.5.  In the NIR,
the H$_2$O-K2 index of \cite{rojas-ayala:2012} provides a spectral
type of M0.0$\pm$0.5. We combine
the optical and infrared types and adopt a final spectral type of
M0.0$\pm$0.5V. Our fundamental parameter and spectral type analyses yield 
consistent results and are also consistent with our visual comparison of \ktwoid 
to M dwarf standards. We compute a distance modulus, $\mu = 4.05\pm0.19$, to \ktwoid by comparing the observed K-band magnitude to tabulated $M_K$ from \cite{pecaut:2013}, which gives a distance of \distepic~pc.

Our HIRES spectrum provides access to the H$\alpha$ line at 6563~\AA. Lines in the
hydrogen Balmer series are associated with magnetic activity in late-type stars and
emission in the H$\alpha$ line is used as a coarse 
age indicator \citep{west:2004,west:2008}.  The HIRES spectrum exhibits H$\alpha$
in absorption, consistent with an inactive star.
We also investigate magnetic activity in \ktwoid by analyzing its UV emission
measured by GALEX \citep{martin:2005}. The star is a weak near-UV (NUV) emitter
and was only marginally detected (2.5$\sigma$) in the far-UV (FUV). Its low NUV
flux and marginal detection in the FUV are consistent with quiescent emission,
similar to other nearby field M dwarfs \citep{shkolnik:2011}. For an M0 dwarf, H$\alpha$ absorption and small UV flux are indicative of weak chromospheric activity and imply an age $\gtrsim$1~Gyr~\citep{west:2008}.

\subsection{Light Curve Fitting}
We analyze the transit light curves using the same approach described by \cite{crossfield:2015}. In brief, we fit each planet's transit
separately using a minimization and Markov Chain Monte Carlo (MCMC) analysis \citep{foreman-mackey:2012}, using JKTEBOP \citep{southworth:2011} to model the light curves.

When modeling the transit photometry, we adopt a linear limb-darkening law. While more complex (e.g. quadratic) limb-darkening prescriptions yield better fits to high SNR light curves, they are overkill given the shallow depths of the \ktwoid transits. We impose Gaussian priors in our analysis. For the limb-darkening parameter, $u$, we assume a distribution with center 0.56 and dispersion 0.13 by referring to the values tabulated by \cite{claret:2012}. All of the MCMC parameters show unimodal distributions, and the inferred parameters are consistent with planets on circular orbits \citep{dawson:2012}.  

Figure~\ref{fig:photometry-fits} shows the resulting photometry and best-fit models, and Table~\ref{tab:planet} summarizes the final values and uncertainties. Of particular interest are the small sizes and low equilibrium temperatures of \ktwoidb~and c. \ktwoidb~and c have \Rp~=~\Rpb~\Re and \Rp~=~\Rpc~\Re, respectively. We estimate equilibrium temperature assuming zero albedo according to the following formula:
\[
\mathrm{T}_{\mathrm{eq}} = \left(\frac{\Sinc}{4\sigma}\right)^{\frac{1}{4}},
\]
where $\sigma$ is the Stefan-Boltzmann constant and \Sinc is the incident stellar flux received by the planet. Planets b and c have equilibrium temperatures of \teqepicb~K and \teqepicc~K, respectively. These planets occupy a domain of planet size and incident stellar radiation which is largely devoid of RV mass measurements and transmission spectra. In Section~\ref{sec:conclusions}, we place the \ktwoid system in the context of the current sample of small transiting planets. 

The transit profile constrains mean stellar density if one assumes a circular orbit. Since we fit each planet separately, we obtain two independent measurements for \rhostarcirc, \rhostarcircb~\gcc and \rhostarcircc~\gcc, which are consistent with \rhostar~=~\rhostarspex~\gcc, derived from spectroscopy. The transit-derived stellar densities are also consistent with each other at the 1-$\sigma$ level, as expected for planets transiting the same star. 

{\renewcommand{\arraystretch}{1.5}
\begin{deluxetable*}{l l l l l}[bt]
\tabletypesize{\scriptsize}
\tablecaption{  Planet Parameters \label{tab:planet}}
\tablewidth{0pt}
\tablehead{
\colhead{Parameter} & \colhead{Units} & \colhead{b} & \colhead{c} 
}
\startdata
   $T_{0}$ & BJD$_\mathrm{TDB} - 2454833$ & $2156.4239^{+0.0025}_{-0.0020}$ & $2155.4708^{+0.0021}_{-0.0017}$\\
       $P$ &          d & $9.32414^{+0.00059}_{-0.00063}$ & $15.50120^{+0.00093}_{-0.00099}$\\
       $i$ &        deg & $88.3^{+1.3}_{-1.1}$            & $89.08^{+0.63}_{-0.75}$\\
 $R_P/R_*$ &         \% & $2.60^{+0.12}_{-0.15}$          & $3.15^{+0.22}_{-0.17}$\\
  $T_{14}$ &         hr & $2.305^{+0.118}_{-0.064}$       & $2.296^{+0.137}_{-0.058}$\\
  $T_{23}$ &         hr & $2.08^{+0.11}_{-0.13}$          & $2.054^{+0.099}_{-0.174}$\\
   $R_*/a$ &         -- & $0.043^{+0.015}_{-0.011}$       & $0.0244^{+0.0100}_{-0.0050}$\\
       $b$ &         -- & $0.70^{+0.15}_{-0.45}$          & $0.66^{+0.19}_{-0.40}$\\
       $u$ &         -- & $0.45^{+0.13}_{-0.11}$          & $0.51^{+0.13}_{-0.11}$\\
\rhostarcirc &     \gcc & $2.7^{+3.7}_{-1.6}$            & $5.4^{+5.4}_{-3.5}$\\
       $a$ &         AU & $0.0731^{+0.0057}_{-0.0067}$    & $0.1026^{+0.0079}_{-0.0094}$\\
       \Rp &        \Re & $1.59^{+0.42}_{-0.44}$          & $1.92^{+0.54}_{-0.52}$\\
     \Sinc &        \Se & $11.0^{+10.1}_{-6.0}$           & $5.6^{+5.1}_{-3.1}$\\
 \teq      &          K & \teqepicb                       & \teqepicc
\enddata
\end{deluxetable*}
}

\subsection{Ruling Out False Positives}
\label{sec:fp}
Thus far, we have explained the transits seen in \ktwoid photometry with two planets around a single M0 dwarf. Here, we assess the possibility that other ``false positive'' interpretations can explain the photometry. One such explanation is that the transits are due to background eclipsing binaries where the eclipses are diluted by \ktwoid. Since an M6 dwarf is $\approx1000\times$ fainter than an M0 dwarf in the \Kepler band ($\Delta M_V = 16.6 - 9.2 = 7.4$),\footnote{Stellar parameters used in this section are drawn from \cite{pecaut:2013}} a pair of maximally eclipsing M6 dwarfs would be diluted to 500~ppm combined with the light of \ktwoid. Thus, we consider false positive scenarios involving M0--M6 companions.

Experience from the \Kepler prime mission has shown that most multi-transiting systems (multis) are bona fide planets. Binary star systems have uniform inclination distributions and are uniformly distributed on the sky. Thus, the probability that two are inclined as seen from Earth and fall in the same software aperture is small. \cite{lissauer:2012} and \cite{rowe:2014} quantified the false positive probability for \Kepler prime multis as < 1\%. This probability represents an upper limit for Campaign 3 multis, given that the field was out of the plane of the Galaxy ($b = -57\deg$), whereas the original \Kepler field was near the plane of the Galaxy, where the density of background stars is higher.

Despite the low probability of the transits being due to background sources, we searched for companion stars falling within the software aperture shown in Figure~\ref{fig:dss}. No such companions separated by more than 5 arcsec are visible in the archival DSS images of \ktwoid. Our NIRC2 image (Figure~\ref{fig:keck}) rules out companions brighter than $\Delta K = 4$ outside of 0.15 arcsec. Over the 61 years between 1954 and 2015, \ktwoid moved by $\sim$5~arcsec relative to distant background stars. Any background stars having $\Delta V$ > 3 within 0.15 arcsec of the current position of \ktwoid would be detectable in the POSS1 plates. The low statistical probability of eclipsing binaries mimicking a multi-transiting system, combined with the non-detection of any background object within our software aperture strongly suggests that the observed transits are not due to background eclipsing binaries.

We now consider the possibility that \ktwoid has a bound companion with its own set of transiting companions. An M6 dwarf is 4 magnitudes fainter than an M0 in K-band. Consulting the contrast curve shown in Figure~\ref{fig:keck}, we rule out companions M6 and earlier separated by more than 0.15 arcsec or 10~AU at a distance of \distepic~pc. 

We searched for close companions using our HIRES spectrum of \ktwoid. Adopting the methodology in \cite{kolbl:2015}, we searched for spectroscopic binaries in our HIRES spectrum. We detect no secondary set of lines from a star having $\Delta V < 4$ mag shifted by more than 15 km/s relative to the lines of the primary star. Shifts of $\Delta v$~>~15~km/s correspond to orbital separations of $\lesssim2$~AU. Thus, the HIRES spectrum rules out bound dwarfs M4 and earlier within 2~AU.

While we have ruled out most of the parameter space where a companion star could be lurking, our search was not exhaustive. However, many of the remaining scenarios involving companions with intermediate separations do not pass stability considerations. A convenient length scale for considering dynamical interactions between planets is the mutual Hill radius,

\[
\RH = \left[\frac{\Min + \Mout}{3 \Mstar} \right]^{1/3} \frac{\ain + \aout}{2},
\]
where $M$ and $a$ denote mass and semi-major axis, respectively. The subscripts ``in'' and ``out'' correspond to the inner and outer planets, respectively. The separation between planet may be expressed in terms of their mutual Hill radius, 
\[
\Delta = \left(\aout - \ain \right) / \RH.
\]
\cite{Gladman93} showed that two planets on initially circular orbits are unstable if $\Delta < 2\sqrt{3} \approx 3.5$.
 
Consider the following false positive scenario: \ktwoid has a bound M5 companion at $a = 5$~AU that hosts its own two-planet system. An M5 dwarf is 5.7 magnitudes fainter than a M0 in $V$-band. In order to reproduce the observed transit depths, the transits across the M5 dwarf must be 190$\times$ deeper to account for dilution. The implied planet radii would much larger: 11~\Re and 14~\Re for planets b and c, respectively. In addition to the fact that M dwarfs rarely host planets planets larger than 3~\Re \citep{Dressing2013}, this hypothetical system is dynamically unstable. Such large radii imply planet masses of > 100~\Me and $\Delta$ < 3.37. Such a system would not be stable over Gyr timescales. We conclude that the interpretation that \ktwoid is a single star orbited by two planets is much more likely than interpretations involving companions that have evaded our follow up imaging and spectroscopy.

\section{Discussion}
\label{sec:conclusions}
\subsection{Planet Masses and Radii at Low Stellar Irradiation}
\label{sec:mass-radii}
\begin{figure}
\centering
\includegraphics[width=0.45\textwidth]{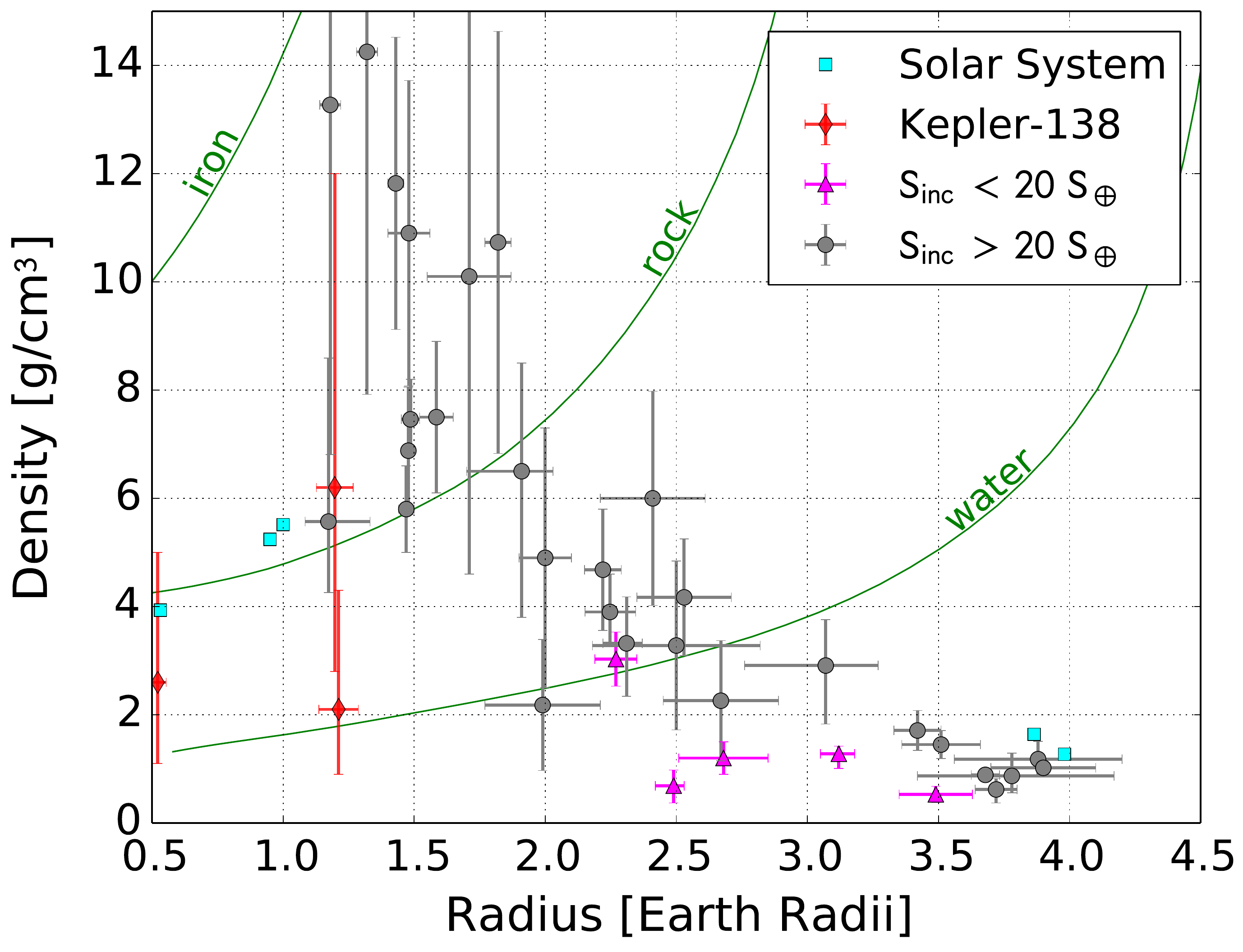}
\caption{Density-radius diagram for all planets with 2-$\sigma$ mass and radius measurements, adopted from the NASA Exoplanet Archive.  Planets with incident flux \Sinc~<~20~\Se (magenta triangles), including Kepler-138b, c, and d (red diamonds), are systematically less dense than those with \Sinc~>~20~\Se (gray circles). Solar system planets are indicated as cyan squares.  The green curves indicate expected planet mass-radius curves for pure iron, water, and rock compositions according to models by \citet{Zeng13}. Below $\sim$1.5~\Re, planets are consistent with primarily rocky compositions. A transition occurs at $\sim$1.5--2.0~\Re and density begins to decrease with radius, which has been interpreted as the onset of substantial atmospheric accretion \citep{marcy:2014,weiss:2014,rogers:2015}. TTV and RV mass measurements of \ktwoidb and c, will help populate the 1.5--2.0 \Re transition region with low \Sinc planets. Such measurements will help test whether low-density outliers like Kepler-138d are the result of limited photoevaporation, distinct formation histories of compact multiplanet systems, or systematic offsets in masses measured with TTVs vs. RVs.}
\label{fig:rho_vs_Rp} 
\end{figure}

We report two small planets with sizes \Rpb~\Re and \Rpc~\Re orbiting \ktwoid, a nearby M dwarf. Figure \ref{fig:rho_vs_Rp} shows a density-radius diagram of all planets with 2-$\sigma$ mass and radius measurements.\footnote{Data are drawn from the NASA Exoplanet Archive, July 28, 2015, http://exoplanetarchive.ipac.caltech.edu} Planets transition from rocky compositions to possessing a significant low-density volatile envelope when their sizes exceed 1.5~\Re \citep{marcy:2014,weiss:2014,rogers:2015}.  While the most likely radii of the \ktwoid planets are larger than 1.5~\Re, their 1-$\sigma$ confidence intervals straddle 1.5~\Re and offer a valuable probe of the densities and bulk compositions near the transition.   

We can estimate planet masses using mass-radius studies from the prime \Kepler mission. Adopting the empirical average mass-radius relationship from \cite{weiss:2014}, we estimate \ktwoidb~and c have masses of \Mp~$\approx$~\Mb~\Me and \Mp~$\approx$~\Mc~\Me, respectively. We emphasize that these mass estimates are crude given the large uncertanties in planet size and the large observed scatter in exoplanet densities for planets having \Rp~=~1.5--2.0~\Re. Assuming circular orbits, these planets would have RV semi-amplitudes $K$ = \kb~\ms and \kc~\ms, respectively. Given the brightness of the host star (\kepmag~=~12.3), measuring planet masses with ground-based spectrometers like HIRES and HARPS is possible, albeit challenging, assuming that the star has low levels of jitter. As a point of reference, \cite{howard:2013} and \cite{pepe:2013} measured the mass of Kepler-78b (\kepmag = 11.6), where $K = 1.66\pm0.41$~\ms.

In addition to their small size, the \ktwoid planets are also noteworthy for their relatively low levels of incident stellar flux.  The radii of close-in planets are likely to be actively sculpted by atmospheric mass loss.  More than 80\% of known planets with \Rp $<$ 2~\Re receive $>$ 100 times more incident stellar radiation than Earth (i.e. \Sinc~>~100~\Se).  Most of the exceptions orbit faint stars for which RV mass measurements and atmospheric characterization is not feasible.  \ktwoid is the brightest star known to host a planet \Rp $<$ 2.0 \Re, \Sinc~<~20~\Se, and $P < 10$ days. It joins Kepler-138 and K2-3 on the list of bright (J $<$ 10.5 mag) stars known to host multiple small planets with low incident flux (\Sinc~<~20~\Se). The \ktwoid system is a laboratory to study the radii of planets where mass loss plays a weaker role.  

TTV measurements of the Kepler-138 system by \citet{Hutter15} showed that at least one of these planets (d) is of surprisingly low density, defying the trend of otherwise rocky planets smaller than 1.5~\Re.  It is unclear whether the low measured density of Kepler-138 d is a product of limited photoevaporation, the result of different formation histories of compact multi-planet systems, or a systematic underestimation of planet masses derived by TTVs \citep[see, e.g.,][]{weiss:2014}.  Combined TTV, RV, and atmospheric follow-up of the \ktwoid system has the potential to determine if small, cool, compact planets are typically low-density, informing models of planet formation and evolution.

\subsection{Transit Timing Variations}
Transit timing variations (TTVs) provide another avenue by which to constrain planet properties. Measuring TTVs with \ktwo is more challenging compared to the prime \Kepler mission because the time baseline over which TTVs can accumulate is much shorter ($\approx$65~days as opposed to $\approx$4~years). Despite these challenges, \cite{Armstrong15} measured TTVs for K2-19b and c, using photometry from \ktwo and the ground-based NITES telescope.

For the \ktwoid system, the ratio of the mean orbital periods is $P_c/P_b$~=~1.6624, close to a 5:3 period commensurability. This proximity indicates that these planets, though low in mass, could interact strongly enough gravitationally to produce observable TTVs. Although a formula for the TTVs of planets near a second order resonance has not yet been derived in the literature, we can use our understanding of TTVs for pairs of planets near first order mean motion resonances to hypothesize the important parameters. The ``super-period'' of the TTV signal, assuming the pair is not dynamically in the 5:3 resonance, will be given by $1/(5/P_c-3/P_b)\approx 1,230$ days. The 65-day \ktwo time series represents only 7\% of this super-period.

The amplitude of the TTV signal will depend linearly on the mass of the perturbing planet, relative to the mass of the host star \citep{Agol} but also on the eccentricities ($e$) and longitudes of pericenter ($\varpi$) of each planet.  For first order resonances, the TTV amplitude and phase depend primarily on the quantities $e_b \cos{\varpi_b}-e_c\cos{\varpi_c}$ and  $e_b \sin{\varpi_b}-e_c\sin{\varpi_c}$ and {\it not} on each of the eccentricities and pericenters individually \citep{Lithwick}. A similar result has been found empirically for second order resonances by fitting the TTVs of Kepler-138. Kepler-138c and d are near the 5:3 resonance and have poorly constrained absolute eccentricities and pericenters, though the differences  $e_b \cos{\varpi_b}-e_c\cos{\varpi_c}$ and  $e_b \sin{\varpi_b}-e_c\sin{\varpi_c}$ are measured well \citep{Hutter15}. Therefore, for second order resonances, these may also be the quantities that control the TTV amplitude.

To estimate the amplitude of the TTVs, we assume nominal masses of \Mp~=~\Mb~\Me and \Mc~\Me. We choose low but reasonable eccentricities of $e_b=e_c = 0.015$. Based on the above reasoning, the maximum TTV signal should be achieved when the pericenters are anti-aligned, while it should be near minimized when the pericenters are aligned. 

\begin{figure}
\begin{center}
	\includegraphics[width=3.3in]{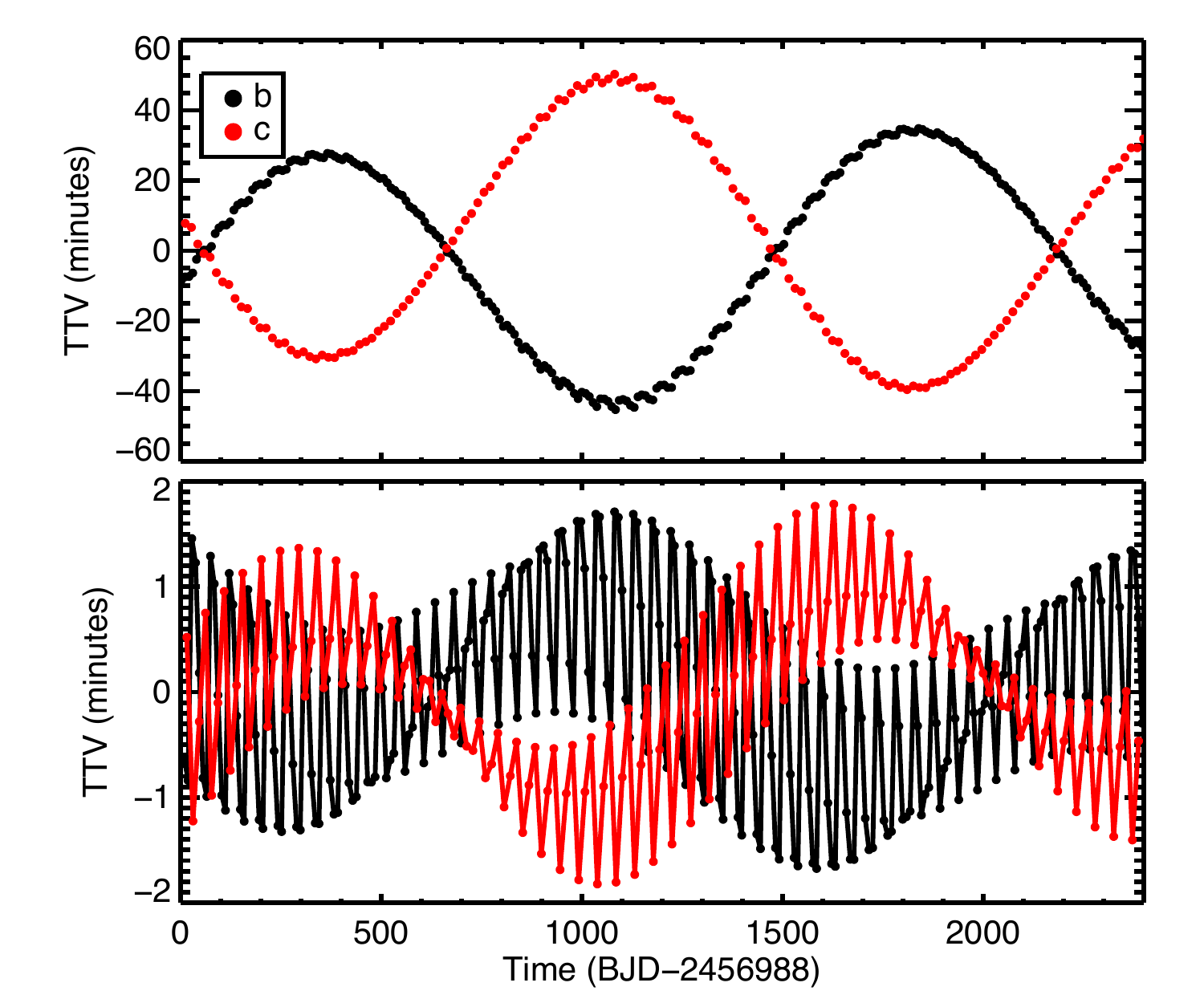}	
\caption{TTV predictions for the two planets based on plausible masses and orbital elements. The case where the pericenters are anti-aligned (top) corresponds to a maximum TTV amplitude, for given masses, while the aligned case (bottom) corresponds to a minimum TTV amplitude. The integration lasts $\approx 2$ ``super periods.''  Eccentricities of 0.015 are assumed. The TTVs (filled circles) are connected in the bottom panel for clarity.}
\label{fig:TTV_predictions}
\end{center}
\end{figure}

In Figure \ref{fig:TTV_predictions}, we show the predicted TTVs in these two limiting cases. These TTVs were calculated with the publicly available n-planet numerical integration and transit timing code TTVFast \citep{TTVFast}. In the bottom panel, the TTVs are dominated by the synodic ``chopping'' signal associated with planetary conjunctions  \citep{NesVor,DeckAgol}. This chopping pattern oscillates on the resonant timescale because of an aliasing effect \citep{DeckAgol}. 

Assuming nominal masses and eccentricities, TTVs range from $\approx$1~min to $\approx$1~hr, depending on the angle between planet pericenters. TTVs of $\approx$1~hr should be detectable from the ground. A detection of TTVs would further constrain the orbital parameters of the \ktwoid planets and compliment RV measurements.

\subsection{Formation Scenarios}
\label{sec:formation}

The proximity of \ktwoidb~and c to the 5:3 resonance may contain clues regarding their formation. The fractional deviation of the current period ratio from exact commensurability is $3(P_c/P_b-5/3)/5 = -0.0025$. \ktwoidb~and c may have been in resonance at some point in the past and may still be in resonance today. At this deviation from exact commensurability, the \ktwoid pair would need eccentricities of approximately 0.05 to be locked in the resonance, according to the estimate of the width of the resonance derived by \citet{Veras} and given in Equation 4.46 of that work. 

An open question is whether the close-in, small planets shown to be common around G, K, \& M dwarfs form {\em in situ} \citep{hansen:2014,chiang:2013} or migrate inward \citep{Izidoro2014,Schlichting14}. Multi-planet systems migrating inward are thought to get trapped in resonance. However, the distribution of period ratios is largely uniform with an excess of planets just outside of first order resonance and a paucity just inside. Theories invoking inward migration have shown that it is possible for planets to break out of resonance \citep{Goldreich14}.

Does the close proximity of the pair to the 5:3 resonance indicate that this system must have assembled via convergent migration and subsequent capture into resonance? There are a handful of known pairs of low-mass planets with period ratios that are even closer to the nearest low-order commensurability (within a deviation of 0.001, in a fractional sense; \citealt{Fabrycky}). However, as noted by \citet{Fabrycky}, even formation processes which produce a smooth background period ratio distribution will contain some pairs that randomly fall very close to resonance. Though the proximity to resonance may dictate the dynamical evolution of these systems, it is not necessarily a smoking gun for assembly via convergent migration. 

\section{Conclusions}
We report two small planet near mean motion resonance orbiting a bright M0 star observed during the \ktwo mission. This system is a convenient laboratory for studying the bulk compositions and atmospheric properties of small planets with low-equilibrium temperatures. Given that \Kepler is expected to observe at least 14 fields in total, we expect many more exciting systems to emerge from the \ktwo mission.

\acknowledgements 
We thank Yoram Lithwick, Kimberly M. Aller, and Brendan Bowler for helpful conversations that improved the manuscript. We thank Lauren M. Weiss for conducting HIRES observations. Support for this work was provided by NASA through Hubble Fellowship grant HST-HF2-51365.001-A awarded by the Space Telescope Science Institute, which is operated by the Association of Universities for Research in Astronomy, Inc., for NASA, under contract NAS 5-26555. This work made use of the SIMBAD database (operated at CDS, Strasbourg, France), NASA's Astrophysics Data System Bibliographic Services, and data products from the Two Micron All Sky Survey
(2MASS), the APASS database, the SDSS-III project, and the Digitized Sky
Survey. 
Some of the data presented in this paper were obtained from the 
Mikulski Archive for Space Telescopes (MAST). 
STScI is operated by the Association of Universities for Research in Astronomy, Inc., 
under NASA contract NAS5-26555. 
Support for MAST for non-HST data is provided by the 
NASA Office of Space Science via grant NNX09AF08G and by other grants and contracts.
Some of the data presented herein were obtained at the W.~M.~Keck
Observatory (which is operated as a scientific partnership among
Caltech, UC, and NASA) and at the Infrared Telescope Facility (IRTF,
operated by UH under NASA contract NNH14CK55B). The authors wish to
recognize and acknowledge the very significant cultural role and
reverence that the summit of Mauna Kea has always had within the
indigenous Hawaiian community.  We are most fortunate to have the
opportunity to conduct observations from this mountain.

{\it Facility:} \facility{\Kepler}, \facility{\ktwo}, \facility{IRTF (SPEX)}, \facility{Keck-II (NIRC2)}, \facility{Keck-I (HIRES)}

\bibliography{full_article.bib}

\end{document}